\documentclass[12pt,preprint]{aastex}

\shorttitle{Nucleosynthesis in Fast Expansions}
\shortauthors{Jordan and Meyer}
\begin{document}
\title{Nucleosynthesis in Fast Expansions \\ of High-Entropy, Proton Rich Matter}
\author{G. C. Jordan IV and B. S. Meyer}
\affil{Department of Physics and Astronomy, Clemson University,
       Clemson, SC 29634-0978, USA}
\email{gjordan@clemson.edu}
\begin{abstract}
We demonstrate that nucleosynthesis in rapid, high-entropy
expansions of proton-rich matter from high temperature and density
can result in a wider variety of abundance patterns than heretofore appreciated.
In particular, such expansions can produce iron-group nuclides, p-process nuclei, or even
heavy, neutron-rich isotopes.  Such diversity arises because the nucleosynthesis
enters a little explored regime in which the free nucleons are not in equilibrium
with the abundant $^4$He. This allows nuclei significantly heavier than iron to form in the presence of abundant free nucleons
early in the expansion.  As the temperature drops, nucleons increasingly assemble into
$^4$He and heavier nuclei.
If the assembly is efficient, the resulting depletion of free neutrons allows disintegration flows
to drive nuclei back down to iron and nickel.
If this assembly is inefficient, then the large abundance of free nucleons
prevents the disintegration flows and leaves a distribution of heavy nuclei
after reaction freezeout.
For cases in between, an intermediate abundance distribution, enriched
in p-process isotopes, is frozen out.  These last expansions may contribute to the solar
system's supply of the p-process nuclides if mildly proton-rich, high-entropy
matter is ejected from proto-neutron stars winds or other astrophysical sites.  Also significant
is the fact that, because the nucleosynthesis is primary, the signature of this nucleosynthesis
may be evident in metal poor stars.
\end{abstract}
\section{Introduction}

Apart from numerous important studies of explosive hydrogen burning in matter
accreted onto white dwarfs or neutron stars and early attempts to understand
the p-process nuclei in terms of hydrogen burning in supernovae
(\citealp{1957RvMP...29..547B,1975ApJ...202..204A}),
astrophysicists have focused relatively
little attention on explosive nucleosynthesis in proton-rich stellar environments.
There are two principal reasons for this.  First, after hydrogen
burning, mainline stellar evolution proceeds at conditions of equal numbers of
neutrons and protons, or, because of weak interactions, at a slight degree
of neutron richness. Accordingly, if the star subsequently explodes, the resulting nucleosynthesis
typically occurs in a neutron-rich environment.  Only if the exploding matter
has not finished hydrogen burning will the nucleosynthesis typically be proton-rich and
proceed in a series of proton capture and $\beta^+$-decay reactions know as the rp-process
\citep{1981ApJS...45..389W}.  Such nucleosynthesis occurs
when matter accreted onto the surfaces of white dwarfs or neutron stars explodes under
degenerate conditions, giving rise to novae or X-ray bursts.

The second reason is that the nucleosynthesis is thought to be already relatively well understood.
The details of the rp-process have been well-studied (e.g, \citealp{2001PhRvL..86.3471S})
since it was first delineated.
For proton-rich matter that achieves higher temperatures,
it is generally imagined that the proton-rich nucleosynthesis would freezeout from an
equilibrium in which iron-group isotopes would dominate the abundances.  The
underlying equilibrium abundance distribution might be slightly modified by proton captures
at the end of the burning.

The purpose of this letter is to show that explosive nucleosynthesis in proton-rich environments
can be much more complex than previously thought.  While it is indeed the case that the
final abundances for many conditions are dominated by iron-group isotopes, for sufficiently
fast, high-entropy expansions of proton-rich matter, isotopes considerably heavier than iron
can form. These other distributions of nuclei can include interesting quantities of light and heavy p-process isotopes or even
neutron-rich species usually ascribed to the r-process.

We have not identified an astrophysical site in which such fast, high-entropy, proton-rich
expansions may occur. The
setting envisioned, however, is a neutrino-driven wind from a high-mass proto-neutron star early
in its epoch of Kelvin-Helmholtz cooling by neutrino emission.
Some calculations find wind entropies $s/k_B$ of order 100--200 and
expansion timescales (which we here define as the radial
expansion timescale, $\tau = r/v$, where $r$ is the radial coordinate and $v$ is the radial
velocity of a parcel of wind matter) as short as a few milliseconds (e.g, \citealp{2001ApJ...562..887T}).
The $Y_e$, that is, the electron-to-baryon ratio, in these winds is set by the interaction of neutrinos and
antineutrinos with free nucleons.  While the antineutrino spectrum from the neutron star is
considerably harder than the neutrino spectrum at late times, which tends to make the
wind neutron rich, it is possible that the two spectra are more nearly equal
earlier.  This could allow for proton-rich (that is, $Y_e > 0.5$) matter since the lower
mass of the proton would favor the reaction $\nu_e + n \to p + e^-$ over the reaction
${\bar \nu_e} + p \to n + e^+$.

\section{Calculations \label{sec:calc}}

The model of the expansion of matter occurring in the fast winds is based on the previous
fast expansion calculations of~\citet{2002PhRvL..89w1101M} and similarly uses the Clemson
Nucleosynthesis Code~\citep{1995ApJ...449L..55M}
with NACRE~\citep{1999NuPhA.656....3A} and NON-SMOKER~\citep{2000ADNDT..75....1R} rate
compilations. In these calculations,
material expands at constant entropy from high temperature and density.  In the present
parameterization, a fluid element in the wind moves out homologously so that its velocity
$v \propto r$, and $r$ grows exponentially in time with timescale $\tau$.  This means the
density declines exponentially with time on a timescale $\tau/3$.  This approximates
the acceleration phase of the wind (e.g., \citealp{1996ApJ...471..331Q}).

We computed several fast expansions.  The final overproductions are presented in
Table~\ref{tab:over}.  For all calculations, the initial composition of the starting
material was a mixture of protons and neutrons which gave the desired initial value of $Y_e$. This
material was given an initial temperature of $T_9 = T/10^9$ K = 10.0 and an initial
density consistent with the chosen value of entropy. The calculations were halted after the
temperature had dropped below $T_9 = 0.01$, by which time all capture reactions had long since ceased
and nuclei were only beta-decaying back to stability.
The values of $\tau$, $s$, and initial $Y_e$ are used when referencing specific calculations. 

\begin{figure}
\epsscale{.50}
\plotone{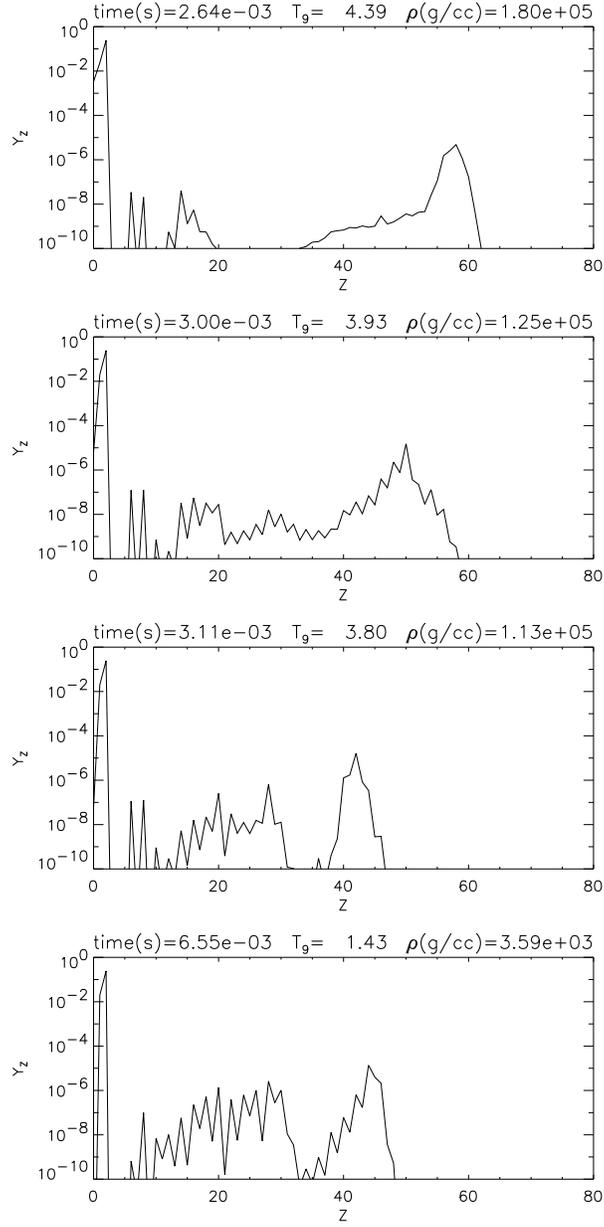}
\caption{Elemental abundance versus mass number $A$ for four times during the
$\tau = 0.003$ s, $s = 145\ k_B$, $Y_e = 0.510$ expansion.\label{fig:fig1}}
\end{figure}

We first focus on a particularly interesting expansion
with $s/k_B = 145$, $\tau = 0.003$ seconds, and $Y_e = 0.510$, which
produces a significant quantity of the light p-process nuclei $^{94}$Mo and $^{96}$Ru.
Figure \ref{fig:fig1}
shows the nuclear abundances as a function of atomic number for four times
in the expansion.  A significant abundance of remarkably high-mass nuclei builds up early in the
expansion.  This surprising result arises from the fact that the abundances of the light
nuclei $^2$H, $^3$H, and $^3$He are low at high entropies.
As a result, nuclear flows assembling $^4$He, which proceed through these
light nuclei, become too slow to maintain an equilibrium between the free nucleons and
$^4$He. As the temperature falls, equilibrium between the free nucleons and $^4$He would increasingly
drive the free nucleons to assemble into $^4$He; however, the slowness of the
requisite reaction flows prevent this from occurring at the necessary rate.
Though this equilibrium fails early (by $T_9 \approx 9$), the system is still primarily composed of $^4$He.
For this particular calculation, at $T_9 \approx 6$ the system is roughly $90\%$ $^4$He,
$6\%$ protons and $4\%$ neutrons by mass with only a slight dusting of heavy nuclei.
By $T_9 \approx 5$ the system is roughly $95\%$ $^4$He, $3.5\%$ protons, and $1.5\%$ neutrons by mass.
From $T_9 = 6$ to $T_9 = 5$, the number of heavy nuclei have increased from $\sim 10^{-8}$
to $10^{-6}$ per nucleon; thus, the heavy nuclei are still negligible in abundance compared
to the protons, neutrons, and $^4$He.
The critical point is that the disequilibrium between free nucleons and $^4$He allows
a large abundance of free nucleons to exist at temperatures that would normally
see the free nucleons locked into $^4$He. The few heavy
nuclei that have been forming at these times thus coexist with a large overabundance of free nucleons
\citep{2002PhRvL..89w1101M}.
Through a rapid sequence of neutron and proton captures these few heavy nuclei are driven to
quite high nuclear mass, as shown in the $T_9 = 4.39$ panel in Figure \ref{fig:fig1}.
The peak at $Z \approx 58$ arises from the fact that the nuclear flows dam up at the closed
neutron shell at $N=82$.

As the temperature continues to fall, new heavy nuclei assemble from the abundant $^4$He nuclei.
The growing abundance of heavy nuclei is then increasingly able to catalyze the synthesis of
$^4$He from free neutrons and protons through reaction cycles such as \\
$^{56}$Ni(n,$\gamma$)$^{57}$Ni(n,$\gamma$)$^{58}$Ni(p,$\gamma$)$^{59}$Cu(p,$\alpha$)$^{56}$Ni.
Once the neutrons disappear (which helped hold the nuclear abundances at the higher mass, nonequilibrium
 distribution), the heavy nuclei begin to disintegrate towards iron-group nuclei, the favored isotopes
in the equilibrium appropriate at those conditions.
The $T_9 = 3.93$ panel of Figure \ref{fig:fig1} shows the abundance distribution shortly after this
disintegration flow begins.  The peak at $Z = 50$ arises from the fact that at this point
the nuclear flow is damming up at the $Z = 50$ closed shell.  By $T_9 = 3.80$, the disintegration
flow has broken through $Z = 50$ and is now dammed up at $N = 50$, and predominantly at the isotope
$^{92}$Mo.  As the temperature continues to decline, proton captures deplete $^{92}$Mo
and create a large abundance of 
$^{94}$Ru and $^{96}$Pd. These isotopes subsequently decay to $^{94}$Mo and $^{96}$Ru after reaction
freezeout.

It is essential to note the large final abundances of protons and $^4$He.  These species, and even
a residual quantity of neutrons, are abundant throughout the disintegration epoch.
This high abundance of neutrons, protons, and $^4$He nuclei hinders the
disintegrations, despite the high temperatures at the time the disintegrations are occurring
($T_9 \approx 4$).  As mentioned above, the high abundance of protons also leads to proton-capture reactions
along the $N = 50$ closed shell that modify the abundances at late times.

Table \ref{tab:over} presents top ten overproduction values from six
calculations of nucleosynthesis in fast, high-entropy, proton-rich expansions.
Calculation 
(B) is the model discussed above ($s = 145\ k_B$, $Y_e = 0.510$, and $\tau = 0.003$ seconds).
As discussed above, the most overproduced isotopes are light p-nuclei, which are made as
$N = 50$ progenitors.  At the top of the list are $^{96}$Ru (made as $^{96}$Pd)
and $^{94}$Mo (made as $^{94}$Ru) at overproduction factors of $\sim 10^6$.
Also significantly produced are $^{95}$Mo (made as $^{95}$Rh) and $^{92}$Mo, which, although
depleted late in the expansion by proton captures, is ultimately produced off of the N=50 shell as $^{92}$Ru.

Calculations (A) and (F) show results for expansions with lower entropy ($s = 140\ k_B$) or
lower $Y_e$ ($Y_e = 0.505$) than in our reference
calculation, (B).  In both cases, the disintegration flow began earlier in the expansion
than in calculation (B) and allowed the nuclei to return to the iron-group isotopes before freezeout.
On the other hand, the larger entropy in calculation (C) delayed the onset of disintegration
and thereby strongly overproduced higher-mass p-process isotopes.  In calculation (D), still
heavier proton-rich isotopes formed.  Rare isotopes such as $^{180}$Ta are particularly
strongly produced in this expansion.  Finally, in calculation
(E), the entropy was sufficiently high that the neutrons disappeared only at low temperature
so the disintegration flow never occurred. This
proton-rich expansion was able to produce heavy, neutron-rich isotopes whose production
is normally associated with rapid neutron capture nucleosynthesis
(cf. \citealp{2002PhRvL..89w1101M}).  These results show
the remarkable sensitivity of the nucleosynthesis to small changes in the expansion parameters.

Because of this sensitivity and the non-equilibrium aspect of the nuclear flows during
key epochs of the expansion, we expect that, for a given set of expansion parameters,
the nucleosynthetic yields of such proton-rich expansions
will be particularly sensitive to nuclear reaction rates on a large suite of isotopes.
This is confirmed in Table~\ref{tab:tab2}, which show results for calculations identical
to calculation (C) but with all charged-particle and electromagnetic rates
on nuclei with $Z \ge 27$ increased or decreased by a factor of two.  In the former case,
the larger reaction cross sections enhance the assembly of nucleons into $^4$He and cause
earlier and more efficient disintegration of the heavy isotopes into iron-group nuclei.
In the latter case, the assembly of nucleons and, hence, the disintegration flows are less
efficient and a heavier distribution of nuclei results than in calculation (C).  Reaction
surveys will be needed to further clarify this issue.

\section{Implications\label{sec:implications}}

As Table~\ref{tab:over} shows, there is a considerable variety of final abundance yields
arising in these expansions.  We do not expect these fast, proton-rich expansions to be
dominant contributors to the solar system's r-process isotopes: the overproductions are low
and the final abundance pattern is different from the solar system's r-process pattern.
On the other hand, these expansions can make interesting quantities of rare isotopes (such
as $^{180}$Ta) and p-process nuclei, perhaps most intriguingly, the light p-process isotopes.

The precise mechanism for the
production of the light p-process nuclei, $^{92,94}$Mo and $^{96,98}$Ru, remains mysterious.
While the heavier p-process isotopes are well accounted for by the ``gamma-process'' in either
core-collapse supernovae \citep{1978ApJS...36..285W} or perhaps in thermonuclear
explosions \citep{1991ApJ...373L...5H}, the same nucleosynthesis underproduces the light
p-isotopes. This is because, unlike the heavy p-process nuclei, the light
p-process isotopes are nearly as abundant as their r-process and s-process counterparts,
which serve as seeds for the gamma-process.
This suggests some other process than the ``gamma-process'' may be responsible for their
origin (although questions remain about the need for an exotic site \citep{2000A&A...358L..67C}).
Suggested other sites or processes are thermonuclear supernovae \citep{1991ApJ...373L...5H},
alpha-rich freezeouts in core-collapse supernovae \citep{1995ApJ...453..792F,1996ApJ...460..478H},
or the rp-process (e.g., \citealp{2001PhRvL..86.3471S}).  Each of these processes or sites
has its own difficulties in either producing the right isotopes or actually ejecting them
into interstellar space, and the question of the site of origin of the light p-process nuclei
remains.

As evident from Table \ref{tab:over}, the typical overproduction factors for calculations that
produce light p-process isotopes are $\sim 10^6$.
Core-collapse supernovae are responsible for production of $^{16}$O, and such supernovae
typically overproduce this isotope by a factor of about 10 (e.g., \citealp{1995ApJS..101..181W}).
If we assume light p-isotopes are 1) produced in rapid, high-entropy, proton-rich
expansions in all core-collapse supernovae,
2) are overproduced at a level of $\sim 10^6$, and 3) are diluted into 10 $M_\odot$ of
ejecta, then we find each supernova must eject $\sim 10^{-4}\ M_\odot$ of this
high-entropy, proton-rich matter.  This is comparable to the estimates of the total mass ejected
in proto-neutron star winds (e.g., \citealp{2001ApJ...562..887T}); therefore, such winds
may contribute interesting amounts of light p-process nuclei to the solar system if they indeed
eject mildly proton-rich matter with the right timescales and entropies.  On the other hand,
some other as yet uncharacterized astrophysical site may achieve the necessary conditions.
A full mapping of parameter space is clearly needed 
to understand this nucleosynthesis process as well as an exploration of its possible astrophysical
settings.

Because winds from proto-neutron stars
stars may evolve from slightly proton rich to neutron rich with time, p-process and r-process
nucleosynthesis may occur in the same site with the former preceding the latter by a few
tenths of a second, the timescale on which the neutrino and antineutrino spectra are
changing.  This may have interesting implications for the coupling
or decoupling of r-process and p-process isotopes in cosmochemical samples (e.g., \citealp{2002Natur.415..881Y}).

Finally, we note that comparison of observations of abundances of Sr, Y, and Zr in low-metallicity stars and
models of the chemical evolution of the
Galaxy hint at a ``lighter element primary process'' (LEPP) that may contribute to the
synthesis of $A \approx 90$ nuclei \citep{2004ApJ...601..864T}.
The process we describe is primary and could therefore be an interesting component of
this inferred LEPP.

\acknowledgments 
The authors are grateful to D. D. Clayton, M. D. Leising, and J. R. King for discussions and to
L.-S. The for assistance. This work was supported by NSF grant
AST 98-19877, NASA grant NAG5-10454, and a grant from DOE's SciDAC programs.

\bibliographystyle{apj}
\bibliography{apj-jour,clemson}

\begin{thebibliography}{18}
\expandafter\ifx\csname natexlab\endcsname\relax\def\natexlab#1{#1}\fi

\bibitem[{{Angulo} {et~al.}(1999){Angulo}, {Arnould}, {Rayet}, {Descouvemont},
  {Baye}, {Leclercq-Willain}, {Coc}, {Barhoumi}, {Aguer}, {Rolfs}, {Kunz},
  {Hammer}, {Mayer}, {Paradellis}, {Kossionides}, {Chronidou}, {Spyrou},
  {degl'Innocenti}, {Fiorentini}, {Ricci}, {Zavatarelli}, {Providencia},
  {Wolters}, {Soares}, {Grama}, {Rahighi}, {Shotter}, \& {Lamehi
  Rachti}}]{1999NuPhA.656....3A}
{Angulo}, C., {Arnould}, M., {Rayet}, M., {Descouvemont}, P., {Baye}, D.,
  {Leclercq-Willain}, C., {Coc}, A., {Barhoumi}, S., {Aguer}, P., {Rolfs}, C.,
  {Kunz}, R., {Hammer}, J.~W., {Mayer}, A., {Paradellis}, T., {Kossionides},
  S., {Chronidou}, C., {Spyrou}, K., {degl'Innocenti}, S., {Fiorentini}, G.,
  {Ricci}, B., {Zavatarelli}, S., {Providencia}, C., {Wolters}, H., {Soares},
  J., {Grama}, C., {Rahighi}, J., {Shotter}, A., \& {Lamehi Rachti}, M. 1999,
  Nuclear Physics A, 656, 3

\bibitem[{{Audouze} \& {Truran}(1975)}]{1975ApJ...202..204A}
{Audouze}, J. \& {Truran}, J.~W. 1975, \apj, 202, 204

\bibitem[{{Burbidge} {et~al.}(1957){Burbidge}, {Burbidge}, {Fowler}, \&
  {Hoyle}}]{1957RvMP...29..547B}
{Burbidge}, E.~M., {Burbidge}, G.~R., {Fowler}, W.~A., \& {Hoyle}, F. 1957,
  Rev. Mod. Phys., 29, 547

\bibitem[{{Costa} {et~al.}(2000){Costa}, {Rayet}, {Zappal{\` a}}, \&
  {Arnould}}]{2000A&A...358L..67C}
{Costa}, V., {Rayet}, M., {Zappal{\` a}}, R.~A., \& {Arnould}, M. 2000, \aap,
  358, L67

\bibitem[{{Fuller} \& {Meyer}(1995)}]{1995ApJ...453..792F}
{Fuller}, G.~M. \& {Meyer}, B.~S. 1995, \apj, 453, 792

\bibitem[{{Hoffman} {et~al.}(1996){Hoffman}, {Woosley}, {Fuller}, \&
  {Meyer}}]{1996ApJ...460..478H}
{Hoffman}, R.~D., {Woosley}, S.~E., {Fuller}, G.~M., \& {Meyer}, B.~S. 1996,
  \apj, 460, 478

\bibitem[{{Howard} {et~al.}(1991){Howard}, {Meyer}, \&
  {Woosley}}]{1991ApJ...373L...5H}
{Howard}, W.~M., {Meyer}, B.~S., \& {Woosley}, S.~E. 1991, \apjl, 373, L5

\bibitem[{{Meyer}(1995)}]{1995ApJ...449L..55M}
{Meyer}, B.~S. 1995, \apjl, 449, L55

\bibitem[{{Meyer}(2002)}]{2002PhRvL..89w1101M}
---. 2002, Physical Review Letters, 89, 231101

\bibitem[{{Qian} \& {Woosley}(1996)}]{1996ApJ...471..331Q}
{Qian}, Y.-Z. \& {Woosley}, S.~E. 1996, \apj, 471, 331

\bibitem[{{Rauscher} \& {Thielemann}(2000)}]{2000ADNDT..75....1R}
{Rauscher}, T. \& {Thielemann}, F. 2000, Atom.~Dat.~Nucl.~Dat.~Tables, 75, 1

\bibitem[{{Schatz} {et~al.}(2001){Schatz}, {Aprahamian}, {Barnard}, {Bildsten},
  {Cumming}, {Ouellette}, {Rauscher}, {Thielemann}, \&
  {Wiescher}}]{2001PhRvL..86.3471S}
{Schatz}, H., {Aprahamian}, A., {Barnard}, V., {Bildsten}, L., {Cumming}, A.,
  {Ouellette}, M., {Rauscher}, T., {Thielemann}, F.-K., \& {Wiescher}, M. 2001,
  Physical Review Letters, 86, 3471

\bibitem[{{Thompson} {et~al.}(2001){Thompson}, {Burrows}, \&
  {Meyer}}]{2001ApJ...562..887T}
{Thompson}, T.~A., {Burrows}, A., \& {Meyer}, B.~S. 2001, \apj, 562, 887

\bibitem[{{Travaglio} {et~al.}(2004){Travaglio}, {Gallino}, {Arnone}, {Cowan},
  {Jordan}, \& {Sneden}}]{2004ApJ...601..864T}
{Travaglio}, C., {Gallino}, R., {Arnone}, E., {Cowan}, J., {Jordan}, F., \&
  {Sneden}, C. 2004, \apj, 601, 864

\bibitem[{{Wallace} \& {Woosley}(1981)}]{1981ApJS...45..389W}
{Wallace}, R.~K. \& {Woosley}, S.~E. 1981, \apjs, 45, 389

\bibitem[{{Woosley} \& {Howard}(1978)}]{1978ApJS...36..285W}
{Woosley}, S.~E. \& {Howard}, W.~M. 1978, \apjs, 36, 285

\bibitem[{{Woosley} \& {Weaver}(1995)}]{1995ApJS..101..181W}
{Woosley}, S.~E. \& {Weaver}, T.~A. 1995, \apjs, 101, 181

\bibitem[{{Yin} {et~al.}(2002){Yin}, {Jacobsen}, \&
  {Yamashita}}]{2002Natur.415..881Y}
{Yin}, Q., {Jacobsen}, S.~B., \& {Yamashita}, K. 2002, \nat, 415, 881

\end{thebibliography}
\clearpage

\clearpage
\begin{deluxetable}{ccccccccccccc}
\tablecolumns{13}
\tabletypesize{\scriptsize}
\tablecaption{Ten most overproduced isotopes in calculations of fast, high-entropy, proton-rich expansions\label{tab:over}}
\tablehead{
\colhead{} & \multicolumn{2}{c}{A\tablenotemark{a}} & \multicolumn{2}{c}{B\tablenotemark{b}} &  \multicolumn{2}{c}{C\tablenotemark{c}} & \multicolumn{2}{c}{D\tablenotemark{d}} &  \multicolumn{2}{c}{E\tablenotemark{e}} & \multicolumn{2}{c} {F\tablenotemark{f}}\\
\cline{2-13} \\
\colhead{rank} & \colhead{$^AZ$} & \colhead{$O$} & \colhead{$^AZ$} & \colhead{$O$} & \colhead{$^AZ$} & \colhead{$O$} & \colhead{$^AZ$} & \colhead{$O$} & \colhead{$^AZ$} & \colhead{$O$} & \colhead{$^AZ$} & \colhead{$O$} 
}
\startdata
1 & $^{59}$Co & 	$5.31\cdot 10^{1}$ & 	$^{94}$Mo & 	$2.15\cdot 10^{6}$ & 	$^{112}$Sn & 	$7.91\cdot 10^{6}$ & 	$^{180}$Ta & 	$7.71\cdot 10^{7}$ & 	$^{201}$Hg & 	$7.91\cdot 10^{5}$ & 	$^{59}$Co & 	$2.78\cdot 10^{2}$ \\ 
2 & $^{45}$Sc & 	$4.98\cdot 10^{1}$ & 	$^{96}$Ru & 	$7.74\cdot 10^{5}$ & 	$^{113}$In & 	$3.88\cdot 10^{6}$ & 	$^{181}$Ta & 	$1.27\cdot 10^{6}$ & 	$^{200}$Hg & 	$5.34\cdot 10^{5}$ & 	$^{60}$Ni & 	$3.95\cdot 10^{1}$ \\ 
3 & $^{60}$Ni & 	$4.25\cdot 10^{1}$ & 	$^{95}$Mo & 	$3.87\cdot 10^{5}$ & 	$^{108}$Cd & 	$2.09\cdot 10^{6}$ & 	$^{176}$Lu & 	$7.70\cdot 10^{5}$ & 	$^{204}$Hg & 	$4.28\cdot 10^{5}$ & 	$^{45}$Sc & 	$2.59\cdot 10^{1}$ \\ 
4 & $^{49}$Ti & 	$3.40\cdot 10^{1}$ & 	$^{92}$Mo & 	$5.40\cdot 10^{4}$ & 	$^{110}$Cd & 	$1.03\cdot 10^{6}$ & 	$^{175}$Lu & 	$7.44\cdot 10^{5}$ & 	$^{199}$Hg & 	$4.25\cdot 10^{5}$ & 	$^{49}$Ti & 	$1.94\cdot 10^{1}$ \\ 
5 & $^{63}$Cu & 	$1.64\cdot 10^{1}$ & 	$^{93}$Nb & 	$2.06\cdot 10^{4}$ & 	$^{106}$Cd & 	$8.52\cdot 10^{5}$ & 	$^{174}$Yb & 	$5.31\cdot 10^{5}$ & 	$^{203}$Tl & 	$3.25\cdot 10^{5}$ & 	$^{63}$Cu & 	$1.84\cdot 10^{1}$ \\ 
6 & $^{48}$Ti & 	$1.31\cdot 10^{1}$ & 	$^{84}$Sr & 	$1.22\cdot 10^{4}$ & 	$^{126}$Xe & 	$4.86\cdot 10^{5}$ & 	$^{178}$Hf & 	$5.23\cdot 10^{5}$ & 	$^{202}$Hg & 	$3.18\cdot 10^{5}$ & 	$^{48}$Ti & 	$1.46\cdot 10^{1}$ \\ 
7 & $^{44}$Ca & 	$1.09\cdot 10^{1}$ & 	$^{97}$Mo & 	$1.17\cdot 10^{4}$ & 	$^{114}$Sn & 	$4.33\cdot 10^{5}$ & 	$^{171}$Yb & 	$5.22\cdot 10^{5}$ & 	$^{198}$Pt & 	$1.94\cdot 10^{5}$ & 	$^{44}$Ca & 	$1.17\cdot 10^{1}$ \\ 
8 & $^{43}$Ca & 	$8.28\cdot 10^{0}$ & 	$^{98}$Ru & 	$6.47\cdot 10^{3}$ & 	$^{115}$Sn & 	$3.27\cdot 10^{5}$ & 	$^{182}$W & 	$4.15\cdot 10^{5}$ & 	$^{182}$W & 	$1.74\cdot 10^{5}$ & 	$^{43}$Ca & 	$1.09\cdot 10^{1}$ \\ 
9 & $^{47}$Ti & 	$6.13\cdot 10^{0}$ & 	$^{91}$Zr & 	$4.39\cdot 10^{3}$ & 	$^{124}$Xe & 	$1.59\cdot 10^{5}$ & 	$^{184}$W & 	$3.61\cdot 10^{5}$ & 	$^{184}$W & 	$1.61\cdot 10^{5}$ & 	$^{47}$Ti & 	$6.41\cdot 10^{0}$ \\ 
10 & $^{50}$Cr & 	$6.11\cdot 10^{0}$ & 	$^{90}$Zr & 	$2.37\cdot 10^{3}$ & 	$^{102}$Pd & 	$1.26\cdot 10^{5}$ & 	$^{176}$Hf & 	$3.54\cdot 10^{5}$ & 	$^{179}$Hf & 	$1.55\cdot 10^{5}$ & 	$^{51}$V & 	$6.08\cdot 10^{0}$ \\

\enddata
\tablenotetext{a}{Calculation A: $\tau=0.003$s, s=140, initial Y$_e$=0.510}
\tablenotetext{b}{Calculation B: $\tau=0.003$s, s=145, initial Y$_e$=0.510}
\tablenotetext{c}{Calculation C: $\tau=0.003$s, s=150, initial Y$_e$=0.510}
\tablenotetext{d}{Calculation D: $\tau=0.003$s, s=160, initial Y$_e$=0.510}
\tablenotetext{e}{Calculation E: $\tau=0.003$s, s=170, initial Y$_e$=0.510}
\tablenotetext{e}{Calculation F: $\tau=0.003$s, s=145, initial Y$_e$=0.505}
\end{deluxetable}

\clearpage
\begin{deluxetable}{ccccc}
\tablecolumns{5}
\tabletypesize{\scriptsize}
\tablecaption{Ten most overproduced isotopes in calculations of with modified reaction rates.\label{tab:tab2}}
\tablehead{
\colhead{} & \multicolumn{2}{c}{A\tablenotemark{a}} & \multicolumn{2}{c}{B\tablenotemark{b}}\\
\cline{2-5} \\
\colhead{rank} & \colhead{$^AZ$} & \colhead{$O$} & \colhead{$^AZ$} & \colhead{$O$} 
}
\startdata
1 & $^{45}$Sc	& $4.19\cdot 10^{1}$	& $^{126}$Xe	& $1.66\cdot 10^{7}$ \\
2 & $^{60}$Ni	& $2.96\cdot 10^{1}$	& $^{124}$Xe	& $7.62\cdot 10^{6}$ \\
3 & $^{49}$Ti	& $2.92\cdot 10^{1}$	& $^{112}$Sn	& $4.29\cdot 10^{6}$ \\
4 & $^{59}$Co	& $1.38\cdot 10^{1}$	& $^{132}$Ba	& $3.53\cdot 10^{6}$ \\
5 & $^{63}$Cu	& $1.27\cdot 10^{1}$	& $^{130}$Ba	& $3.53\cdot 10^{6}$ \\
6 & $^{48}$Ti	& $1.20\cdot 10^{1}$	& $^{113}$In	& $2.44\cdot 10^{6}$ \\
7 & $^{44}$Ca	& $9.89\cdot 10^{0}$	& $^{138}$Ce	& $1.96\cdot 10^{6}$ \\
8 & $^{43}$Ca	& $7.42\cdot 10^{0}$	& $^{128}$Xe	& $1.15\cdot 10^{6}$ \\
9 & $^{47}$Ti	& $5.41\cdot 10^{0}$	& $^{114}$Sn	& $9.30\cdot 10^{5}$ \\
10 & $^{50}$Cr	& $4.79\cdot 10^{0}$	& $^{115}$Sn	& $9.07\cdot 10^{5}$ \\
\enddata
\tablenotetext{a}{Calculation A: $\tau=0.003$s, s=150, initial Y$_e$=0.510; charged particle and EM rates for $Z\geq27 \times 2$ }
\tablenotetext{b}{Calculation B: $\tau=0.003$s, s=150, initial Y$_e$=0.510; charged particle and EM rates for $Z\geq27 \times 0.5$}
\end{deluxetable}
\end{document}